# Using Temporal Data for Making Recommendations

Andrew Zimdars, David Maxwell Chickering, Christopher Meek

## Abstract

We treat collaborative filtering as a univariate time series problem: given a user's previous votes, predict the next vote. We describe two families of methods for transforming data to encode time order in ways amenable to off-the-shelf classification and density estimation tools. Using a decision-tree learning tool and two real-world data sets, we compare the results of these approaches to the results of collaborative filtering without ordering information. The improvements in both predictive accuracy and in recommendation quality that we realize advocate the use of predictive algorithms exploiting the temporal order of data.

## 1 Introduction

The collaborative filtering problem arose in response to the availability of large volumes of information to a variety of users. Such information delivery mechanisms as Usenet and online catalogs have created large stores of data, and it has become the users' task to discover the most relevant items in those stores. Rather than requiring that users manually sift through the full space of available items, trusting that authors respect the available system of topics, CF tools recommend items of immediate or future interest based on all users' expressed preferences ("votes"), suggesting those items of interest to other users with similar tastes. These votes may be either explicit, as in response to a direct inquiry, or implicit, as by the choice to follow one hyperlink instead of others.

In general, algorithms for the CF task, such as those explored by Breese, Heckerman and Kadie (1998), have not relied on the order in which users express their preferences. Vector-space methods draw heavily on work in the information retrieval literature (see, e.g., Baeza-Yates and Ribeiro-Neto, 1999), where individual documents are treated as a "bag of words". Likewise, probabilistic techniques (e.g. Hofmann and Puzicha, 1999 and Heckerman, Chickering, Meek, Rounthwaite and Kadie, 2000) have computed probability distributions over recommendations conditioned on the entire vote history without regard to time order. In the CF literature, a "bag of votes" (i.e. atemporal) assumption prevails, and the collaborative filtering problem is cast as classification (with classes "relevant" and "irrelevant") or density estimation (of the probability that a document is relevant, given a user's votes).

We instead consider collaborative filtering as a univariate time series prediction problem, and represent the time order of a user's votes explicitly when learning a recommendation model. Further, we encode time order by transforming the data in such a way that standard *atemporal* learning algorithms can be applied directly to the problem. Other authors (cf. Mozer, 1993) have applied atemporal learning techniques to temporal data; we describe here two successful generic techniques. As a result, researchers can simply transform their data as we describe and apply existing tools, instead of having to re-implement various collaborative filtering algorithms for awareness of vote order. Our approach allows CF models to encode changes in a user's preferences over time. It also allows models to represent (indirectly) structure built into the feature space that would be lost in a bag of votes representation. For example, Web page viewing histories ordered by page request can express the link structure of a Web site because a user is most likely to follow links from his current page. Similarly, television viewing histories encode the weekly schedule of shows: a viewer cannot hop from *Buffy the Vampire Slayer* to *Dawson's Creek* if the two are not contemporaneous.

For simplicity, we assume for the remainder of this paper that user preferences are expressed as implicit votes (see, e.g., Breese et al., 1998). That is, a users' vote history is a list of items that the user preferred, as opposed to an explicit *ranking* of the items. In a movie domain, for example, this means that a user's vote history is simply a list of movies that he watched, and we assume that he preferred those movies to the ones he did not watch. We note, however, that the transformations we describe are easily generalized to explicit voting.

In Section 2, we present two methods for transforming user vote histories that encode time-order information in ways that traditional atemporal modeling algorithms can use. In Section 3, we discuss three candidate models that can be learned from standard algorithms applied to the transformed data. In Section 4, we describe the data sets and criteria by which we will compare our approaches, and in Section 5 we present our experimental results from using decision-tree learning algorithms.

## 2 Data Transformations

In this section, we describe two methods that transform time-ordered vote histories into a representation that traditional atemporal modeling algorithms can use; we call this representation the *case representation*. In the case representation, the data $D$ consists of a set of cases (or *records*) $\{C_1, \ldots, C_m\}$, where each case $C_i = \{x_1, \ldots, x_n\}$ consists of a value for zero or more of the variables in the domain $\mathbf{X} = \{X_1, \ldots, X_n\}$.

The important (sometimes implicit) assumption of modeling algorithms that use the case representation is that the observed cases are independent and identically distributed (iid) from some joint probability distribution $p(X_1, \ldots, X_n)$[1]; an equivalent Bayesian assumption is that the cases are *infinitely exchangeable*, meaning that any permutation of a set of cases has the same probability. The learning algorithms use the observed case values in $D$ to identify various models of the generative distribution.

As an example, consider the problem of predicting whether or not a particular person will watch some television show based on that person's age and gender. Using the case representation, we might assume that *all people* are drawn from some joint probability distribution $p(S, A, G)$, where $S$ is a binary variable that indicates whether or not a person watches the show, $A$ is a continuous variable denoting a person's age, and $G$ is a binary variable that denotes the person's gender. Under the iid assumption a learning algorithm can use observed values of $S$, $A$, and $G$ for other people in the population to estimate the distribution $p(S|A, G)$, then make a prediction about the particular person of interest with that distribution.

In the following sections, we describe how data that contains vote histories can be transformed, using various assumptions, into the case representation so that standard machine-learning algorithms can be used to predict the next vote in a sequence. First, we need some notation.

We use *item* to denote an entity for which users express preferences by voting, and we use $\gamma$ to denote the total number of such items. For example, in a movie-recommendation scenario, $\gamma$ is the total number of movies considered by the collaborative-filtering system. For simplicity we refer to each item by a one-based integer index. That is, the items in the system are mapped to the indices:

$$\{1, \ldots, \gamma\}$$

We use $\mathbf{V}_i$ to denote the $i^{th}$ vote history (i.e. user's votes). In particular, $\mathbf{V}_i$ is an *ordered* list of votes:

$$\{V_i^1, \ldots, V_i^{N_i}\}$$

where $V_i^j$ denotes the item index of the $j^{th}$ vote in the list, and $N_i$ is the total number of votes made by user $i$.

As an example, suppose there are four movies *The Matrix*, *Star Wars*, *A Fish Called Wanda* and *Pulp Fiction* having indices 1,2,3 and 4, respectively. Suppose there are two movie watchers in the domain: User 1 watched *The Matrix* and then watched *Pulp Fiction*, and user 2 watched *Star Wars*, then watched *Pulp Fiction*, and then watched *The Matrix*. Then we would have $\mathbf{V}_1 = \{1, 4\}$ and $\mathbf{V}_2 = \{2, 4, 1\}$.

For each of the transformations below, we show how to convert from a set of vote histories into (1) a set of domain variables $\mathbf{X} = \{X_1, \ldots, X_n\}$, and (2) a set of cases $\{C_1, \ldots, C_m\}$, where each case $C_i$ contains a set of values $\{x_1, \ldots, x_n\}$ for the variables in $\mathbf{X}$. We also describe what assumptions are made in the original domain in order for the resulting cases to be iid.

### 2.1 The "Bag-of-votes" Transformation

The first transformation we consider disregards the order of previous votes, corresponding to the assumption that vote order does not help predict the next vote. As noted above, this "bag-of-votes" approach is the approach taken by many collaborative-filtering learning algorithms.

---

[1] In fact, if we are interested in learning a conditional model for $\mathbf{Y} \subset \mathbf{X}$, we often need only assume that the values for the variables in $\mathbf{Y}$ are independent samples from some $p(\mathbf{Y}|\mathbf{X} \setminus \mathbf{Y})$

For each item $k$, where $1 \leq k \leq \gamma$, there is a binary variable $X_j \in \mathbf{X}$, whose states $x_j^1$ and $x_j^0$ correspond to *preferred* and *not preferred*, respectively. There are no other variables in $\mathbf{X}$. For each vote history $\mathbf{V}_i$, we create a single case $C_i$ with the following values: if item $j$ occurs at least once anywhere in the sequence $\mathbf{V}_i$, then the value $x_j$ in $C_i$ is equal to $x_j^1$. Otherwise, the value of $x_j$ in $C_i$ is equal to $x_j^0$.

The assumption that the cases are iid corresponds to assuming that the (unordered) votes of all vote histories (i.e. users) are all drawn from the same distribution. Under this assumption, we can use an atemporal learning algorithm with the cases from previous vote histories learn a model for $p(X_j|\mathbf{X}\setminus X_j)$ for all $X_j \in \mathbf{X}$, and then use these models to predict the next vote[2] for any vote history.

### 2.2 The Binning Transformation

The second transformation we consider can be helpful when user preferences change over time. Although the transformation does not explicitly use the order of the votes, it can exploit temporal structure. The idea is to (1) separate vote histories into bins by their size, (2) transform the histories from each bin into the case representation using the "bag-of-votes" transformation described above, and (3) learn a separate model from the data in each such bin. When it comes time to predict the next vote in a sequence of size $k$, we use the model that was learned on the cases derived from the vote histories in the bin corresponding to $k$.

Suppose, for example, that we would like to train one or more models in order to recommend movies to people. It might be reasonable to assume that the optimal model for predicting the third movie for someone may not be a very good model for predicting the 100th movie. With binning, we divide up the range of the number of movies that have previously been seen into separate bins, and learn a recommendation model for each. Thus, we might end up with three models: (1) a simple model that predicts popular movies for people who do not go to the movies much, (2) a model that perhaps identifies general viewing preferences (e.g. comedies) for the typical viewer, and (3) a model that identifies subtle preference trends for heavy movie watchers.

In order to perform binning, there are a number of parameters that need to be set. First, we need to decide how many bins to use. Second, we need to decide, for each bin, what history lengths should be included in that bin.

For the experiments that we present in Section 4, we tried both two and four bins. For each bin, we set a minimum and maximum value for the length of the contained histories. We chose this minimum and maximum such that the *total* number of votes in each bin are roughly the same.

As described above, the binning approach assigns each vote history to exactly one bin. An alternative approach, which we call the *prefix approach*, is to allow a single vote history to contribute to multiple bins by adding an appropriate prefix to all of the "previous" bins. As an example, suppose there are three bins that accommodate histories of length up to 5, 10, and 100. In the prefix approach, a vote history of length 90 will have (1) the first five votes added to the first bin, (2) the first ten votes added to the second bin, and (3) the whole history added to the third bin.

The choice of whether or not to use the prefix approach to binning will depend on user behavior and domain structure. We identify the following two hypotheses that can help determine which method is most appropriate.

- The "expert/novice" hypothesis: Users with long vote histories ("experts" in the domain) have fundamentally different preferences than users with short vote histories ("novices"). As a result, we expect that omitting prefixes of longer vote histories from bins for shorter vote histories will result in better predictive accuracy than the prefix approach. The expert/novice hypothesis might hold when predicting preferences for television viewing, where couch potatoes might have different viewing habits than occasional viewers. On a Web site, heavy users tend to navigate very differently than "shallow browsers" (cf. Huberman et al., 1998).

- The "everyone learns" hypothesis: Users with long vote histories once expressed similar preferences to users with short vote histories. Under this hypothesis, we expect that prefixes of long vote histories will be distributed similarly to short vote histories, and therefore their inclusion in the corresponding bins will provide useful data for the model-building algorithm; as a result, we hope that the resulting models will be more accurate. One can also interpret this hypothesis from the perspective of domain structure constraining user behavior. For users of a Web portal, initial votes may be restricted to the home page and top-level categories linked from that page. For subsequent page hits, available links may constrain possible user votes. In this domain, we would expect users to have similarly-distributed vote prefixes because site structure does not allow much room for inno-vation.

---

[2]There are some subtleties, addressed below, about how this prediction is made.

For the domains we consider in Section 4, the latter hypothesis seems more appropriate; although we ideally should have compared the two, in the interest of time we only used the prefix approach in our experiments. We chose the bin boundaries so that the total number of votes of the *original* (i.e. non-prefix) histories in each bin were roughly the same.

Whether or not we use the prefix approach, the additional computational overhead of binning over no binning is proportional to a constant factor (the number of bins), because each bin will contain no more votes, and no more vote histories, than would a single model computed using the entire vote set.

Structural aspects of some prediction domains can make difficult the choice of vote sub-histories to augment data for binning. Web sites tend to have a hierarchical structure with a home page at the root, but the same cannot be said for television programming schedules, which reflect periodic structure. When predicting television viewing habits given a "snapshot" of user viewing histories, prefixes may not reflect the periodic nature of the program schedule. In such domains, different choices of contiguous vote sub-histories may be appropriate, but the resulting profusion of data might render binning impractical.

We should point out that binning can be applied to collaborative filtering problems in which the temporal order of the votes is unknown. Although the prefix approach may not be appropriate, binning based on the number of votes can potentially lead to significantly better accuracy in atemporal domains. Consider, for example, the problem of recommending items in a grocery store based on the products bought (the recommendation may appear as a targeted coupon on a receipt). It might turn out that, regardless of the *order* in which people put groceries in their shopping cart, the *number* of items in their cart may indicate very different shopping behavior; consequently the binning approach might yield significantly better models than a system that ignores the number of votes.

### 2.3 Data Expansion

The final data transformation we consider, which we call *data expansion*, finds inspiration in the language modeling literature (see, e.g., Chen and Goodman, 1996). This method of data expansion distinguishes the most recent $n$ votes from the entire vote history, as well as identifying the order of the most recent votes. All of the variables that we create in the transformation are binary, and have states $x1$ and $x0$ correspond to *preferred* and *not preferred*, respectively.

In the case representation, we create one binary variable for each of the $\gamma$ items in the domain: $\mathbf{X}^T = \{X_1^T, \ldots, X_\gamma^T\}$. The "T" superscript in $X_k^T$ is meant to indicate that this is a "target variable" that represents whether or not the next vote is for item $k$.

The data expansion transformation is parameterized by a *history length* $l$; this parameter, which corresponds to the "$n$" parameter in an $n$-gram language model, determines how far back in the vote history to look when predicting the next vote. For each integer history $1 \leq j \leq l$, we again create one binary variable for each of the $\gamma$ items in the domain: $\{X_1^{-j}, \ldots, X_\gamma^{-j}\}$. The "$-j$" superscript in $X_k^{-j}$ is meant to indicate that this variable represents whether or not $j^{th}$ previous vote (from the one we're predicting) is for item $k$. We use $\mathbf{X}^L$ to denote the set of all lagged variables (e.g. $\{X_1^{-1}, \ldots, X_\gamma^{-1}\}, \{X_1^{-2}, \ldots, X_\gamma^{-2}\}$).

There is a final set of $\gamma$ variables consisting of, for each item, an indicator of whether or not that item was voted for at least once previously in the given vote history. We use $\mathbf{X}^C = \{X_1^C, \ldots, X_\gamma^C\}$ to denote these variables. In language-modeling parlance, these variables are known as *cache* variables.

In contrast to the "bag-of-words" approach, where each vote history was transformed into a single case, in the data expansion transformation, each *vote* in every history gets a corresponding case. In particular, for vote $V_i^j$, which is the $j^{th}$ vote in the $i^{th}$ vote history, we define the values for all of the variables as follows. For simplicity, let $v = V_i^j$. We set the value of target variable $X_v^T$ to $x1$, and we set the value of all other target variables to $x0$. For each history variable $X_k^{-j}$, where $1 \leq j \leq l$, we set the corresponding value to either $x1$ if the $j^{th}$ previous vote in history $i$ has value $k$, or $x0$ otherwise. Finally, we set the value of each cache variable $X_k^C$ to either $x1$ if item $k$ occurs as a vote (at least once) previous to $V_i^j$ in $\mathbf{V}_i$, or $x0$ otherwise.

We should point out that in order to feasibly learn a model using the cases that result from the data-expansion transformation, the learning algorithm(s) need to use a sparse representation for the cases. See (e.g.) Chickering and Heckerman (1999) for a discussion.

Consider our movie example again. For simplicity, we use $M$, $S$, $F$, and $P$ to label all variables we create corresponding to movie items *The Matrix*, *Star Wars*, *A Fish Called Wanda* and *Pulp Fiction*. Furthermore, we use 1 and 0 to denote the values *preferred* and *not preferred*, respectively.

Suppose we want to transform a vote history containing *The Matrix*, *Pulp Fiction*, and *Star Wars*, in that order, into the case representation with a history length of one. First we define the variables

$$\mathbf{X} = \{ \; M^T, S^T, F^T, P^T, \\ M^{-1}, S^{-1}, F^{-1}, P^{-1}, \\ M^C, S^C, F^C, P^C \}$$

Next, we consider each vote in the history, and create a case for each one. Table 1 shows the case values that result.

The learning algorithm we use should build a model for each of the target variables, using all non-target variables as predictor variables. That is, we would like the model to estimate, for each target variable $X_j^T \in \mathbf{X^T}$, the distribution $p(X_j^T|\mathbf{X^L}, \mathbf{X^C})$.

The iid assumption in the case representation—after performing the data-expansion transformation with history-length $l$—implies that each vote is drawn from a distribution that depends on (1) the values of the previous $l$ votes and (2) the presence or absence of at least one vote for previous items.

## 3 Models

In this section, we describe some well-known models that can be used for collaborative filtering applications; when learned from data that is transformed as described in the previous section, these models can exploit the vote order to improve recommendation accuracy.

### 3.1 Memory-based algorithms

Memory-based collaborative filtering algorithms predict the votes of the active user based on some partial information about the active user and a set of weights calculated from the user database. Memory-based algorithms do not provide the probability that the active user will vote for a particular item. Instead, the active user's predicted vote an item is a weighted sum of the votes of the other users. See Breese et. al (1998) for a more detailed discussion.

### 3.2 Cluster models

A standard probabilistic model is the naïve Bayes model with a hidden root node—one where the probabilities of votes are conditionally independent given membership in an unobserved class variable $C$, where $C$ ranges over a fairly small set of discrete values. This corresponds to the intuition that users may be clustered into certain groups expressing common preferences and tastes. The joint probability distribution for this model is expressed as follows:

$$P(C = c, \mathbf{v}_1, \ldots, \mathbf{v}_n) = P(C = c) \prod_{i=1}^{n} P(\mathbf{v}_i \mid C = c) \quad (1)$$

The parameters of this model can be learned using the EM algorithm (see Dempster, Laird and Rubin, 1977). Cheeseman and Stutz (1995) provide details of a specific implementation of the learning algorithm.

In this setting, prediction for collaborative filtering follows from the density estimation problem, as the model predict the item(s) most likely to receive an affirmative vote given the user's vote history.

Other latent class models (Hofmann and Puzicha, 1999) have been proposed for collaborative filtering which place user and item on an equal footing. These permit construction of a two-sided clustering model with preference values, but they depend on multinomial sampling of (user, item) pairs, and as such do not generalize naturally to new users.

### 3.3 Decision-tree models

The approach that has proven most effective in previous work (cf. Heckerman et al., 2000) constructs a forest of probabilistic decision trees, one for each item in the database, using a Bayesian scoring criterion (Chickering, Heckerman, and Meek, 1997). This provides a compact encoding of conditional probabilities of recommendations, given previous votes.[3] We use this approach in Section 4 to evaluate our data transformations.

### 3.4 Alternative models

The data expansion technique discussed in Section 2.3 suggests the application of language-modeling algorithms to collaborative filtering. We have conducted limited experiments with variants of $n$-gram language models, and the results are promising (although we do not present them here).

Hidden Markov models (HMMs) also recommend themselves in this setting, but in our experience they are ill-suited to a naïve representation of the data, where each possible vote corresponds to exactly one feature. This reflects in part the number of parameters that must be estimated when running EM for an HMM: if the model admits $c$ hidden states, then there are $mc + c^2 + c$ parameters to estimate for the posterior probabilities of states, the state transitions, and the state priors. Moreover, models are slow to converge because collaborative filtering data tend to be very sparse, in that few users vote on any one item. As a result, evidence for estimating a particular variable is rarely presented in training. This sparsity is integral to the collaborative filtering problem, but lethal to accurate estimation. Finally, HMMs discard much of a user's history in making predictions, and our experiments indicate that a long history can be informative.

---

[3]It also permits the construction of a family of graphical models known as *dependency networks*, which have expressive strength similar to Markov networks.

Table 1: Case values created for the movie example with the data expansion method.

| Vote | $M^T$ | $S^T$ | $F^T$ | $P^T$ | $M^{-1}$ | $S^{-1}$ | $F^{-1}$ | $P^{-1}$ | $M^C$ | $S^C$ | $F^C$ | $P^C$ |
|---|---|---|---|---|---|---|---|---|---|---|---|---|
| *The Matrix* | 1 | 0 | 0 | 0 | 0 | 0 | 0 | 0 | 0 | 0 | 0 | 0 |
| *Pulp Fiction* | 0 | 0 | 0 | 1 | 1 | 0 | 0 | 0 | 1 | 0 | 0 | 0 |
| *Star Wars* | 0 | 0 | 1 | 0 | 0 | 0 | 0 | 1 | 1 | 0 | 0 | 1 |

## 4 Experiments

In this section, we describe the experiments we performed to demonstrate that using vote order can improve the accuracy of models.

We conducted our experiments using two real-world data sets, both of which are Web user traces. In each, the notion of "user" corresponds to a server session, and a page request was interpreted as an affirmative vote.

The first data set consists of session traces from http://research.microsoft.com/. The training data encompassed 110587 page requests from 27595 users over three days in late August 1999, and the test data included 54843 requests from 13563 users on 14 September of the same year. The requests span a total of 8420 URLs, roughly 400 of which correspond to 404 errors for invalid URLs. The average length of a session trace was 4.007 votes, with a median length of 2, and the longest trace was 93 votes.

The second data set uses session traces from http://www.msnbc.com/, corresponding to an 80%/20% split of users on 22 December 1998. The training data include roughly 1.28 million requests from 475769 users, while the test data include 178158 requests from 87714 users. The requests in these two data sets span 1001 URLs; it is unclear whether any of these represent invalid URLS. The average length of a session trace was 2.696 vote, with a median length of 2 and a longest trace of 407 votes.

Unfortunately, we did not identify other publicly-available data that records user preferences in time order. The authors' experience with other data suggests that the techniques outlined here may prove fruitful with other types of sequential data.

We used probabilistic decision-tree models for our experiments, and compared both binning and data expansion to the default "bag-of-votes" approach of ignoring the data order. For all of the experiments, we learned a single decision tree per page to predict whether the user requests that page, based on the transformed data available at that time. We used a greedy tree-growing algorithm in conjunction with the Bayesian score described by Chickering et. al (1997). In particular, the score evaluated the posterior model probability using a flat parameter prior, and a model prior of the form $\kappa^f$, where $f$ is the number of free parameters in the tree. We used $\kappa = 0.01$ for all of the experiments.

In all of the data transformations described in the previous section, we created a separate binary variable for each item that denoted whether or not the next vote will be for that item. Defining the variables this way can be problematic for any learning algorithm using finite data that does not enforce the constraint that the next vote will be for exactly one item. In particular, the algorithm we used to learn a forest of decision trees did not enforce this constraint. We solved this problem by using the decision trees to calculate the posterior probability that each item would be the next vote, then renormalizing.

We applied two evaluation criteria in our experiments. For all prediction algorithms, we adopted the "CF accuracy" score outlined by Heckerman et al. (2000), and specialized it to compute the CF score with respect to the next item in the user's history only. The CF accuracy score attempts to measure the probability that a user will view a recommendation presented in a ranked list with other recommendations. To approximate this probability, let $p(k) = 2^{-k/\alpha}$ denote the probability that the user views the $k$th item on his list (where $k$ counts from 0). For the experiments presented here, we chose a half-life of $\alpha = 10$. We computed for each user $i$, and for each vote $v_{ij}$ in his vote history, a ranked list of recommendations given $v_{i1}, \ldots, v_{i(j-1)}$.

One may compute the CF accuracy of a general list $L$ of test items spanning $n$ users. Suppose the model recommends $R_i$ items to each user, and the users actually prefer sets of $M_i$ items. Let $\delta_{ik}$ denote the indicator that user $i$ prefers the $k$th recommendation. Then

$$\text{accuracy}_{CF}(L) = \frac{1}{n} \sum_{i=1}^{n} \frac{\sum_{k=0}^{R_i-1} \delta_{ik} p(k)}{\sum_{k=0}^{M_i-1} p(k)} \qquad (2)$$

Let $k_{ij}$ be the ranking assigned by our model to vote $v_{ij}$. Scoring one vote at a time, CF accuracy simplifies to

$$\text{accuracy}_{CF} = \frac{1}{\sum_i N_i} \sum_{i=1}^{n} \sum_{j=1}^{N_i} 2^{1/\alpha} p(k_{ij}) \qquad (3)$$

One may compute CF accuracy for any CF algorithm that generates a ranked list of recommendations, but it provides a criterion specific to the collaborative filtering task. For the probability models we evaluated, we also computed the mean log-probability assigned to each of the user's actual votes, given the preceding vote history. (This log-probability was normalized over all items in dependency-network models to compensate for potential inconsistencies).

Note that CF accuracy is a function of the *relative* magnitude of density estimates, while the log score depends on the *absolute* magnitude of the estimates.

## 5 Results

The results presented below correspond to three families of models. The "Baseline" results derive from a forest of decision trees trained on bag-of-votes data, shown to be a one of the best models for CF (Breese et al., 1998). "2 Bins" and "4 Bins" experiments applied the binning method described in section 2.2. Two or four decision trees are constructed for each Web page, but only one is chosen (according to the partial history at hand) to make a prediction. The "DE-" experiments expand data as in section 2.3, with histories of length 1, 3, and 5.

Figure 1 and Figure 2 show the CF scores and log scores, respectively, for all of the models in the MSNBC domain.

There are some interesting observations to make about these results. First, we see that for the collaborative-filtering score, the score got *worse* as we increased the number of bins. This may be an artifact of the sparsity of long traces in Web surfing data, a phenomenon that has been observed elsewhere (e.g., Huberman et al., 1998). This may not impair work in other domains; our experience with data suggests that other frequency functions for user history length can have thicker tails.

Second, we see that all of the data-expansion models performed significantly better than the baseline with respect to CF accuracy, but that performance did not increase as a function of history length. This might suggest that Web page requests depend more strongly on immediate links than on the short-term history, and that data expansion mainly embodies this structural element of the Web surfing domain. (One should not interpret this as a Markov assumption; in our experience, the cache variables strongly influence prediction.) The higher CF accuracy results suggest that the *relative* magnitude of density estimates is more often accurate for data-expanded models than binned models, and these relative estimates determine which pages show up in a recommendation list.

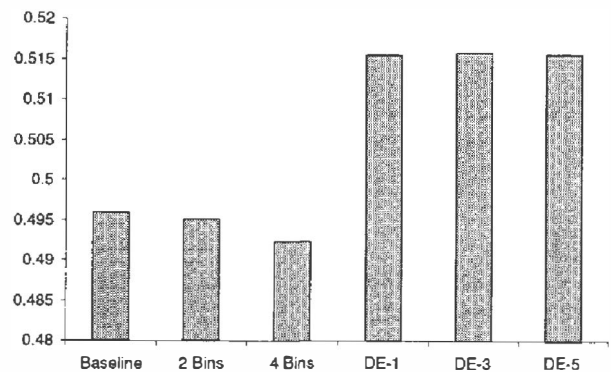

Figure 1: Collaborative filtering scores of the models constructed for the MSNBC domain.

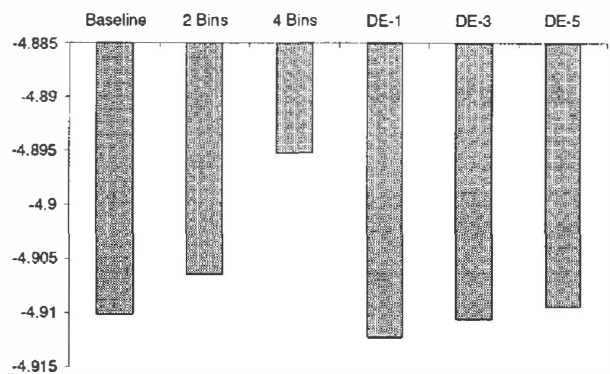

Figure 2: Log-probability scores of the models constructed for the MSNBC domain.

Our results show that unlike for the CF score, the binning approach dominated both the baseline and the data-expansion models for log-probability predictive accuracy. For this score, the data-expansion models improved as the history length increased, but only the model with the longest history (five) was competitive with the baseline model. We suspect that the data were too sparse to permit accurate parameter estimates for the models learned under data expansion. In particular, there were roughly 50 percent more pa-rameters to train in each of the data-expansion models than in the other models, which leads us to suspect that the learning algorithm over-fit for these models to some degree. In retrospect, we regret the choice of a single value of the model-prior parameter $\kappa$ for all data transformations. We expect that if we had tuned this parameter by splitting up the training data and maximizing a hold-out prediction accuracy, we would have identified a smaller $\kappa$ for the data-expansion models that yielded better results for both criteria on the tests set. Improvements in log score as history length increase demonstrate the value of the additional information encoded by the expanded data, which compensates in part for having too few data points per parameter.

Figure 3 and Figure 4 show the CF scores and log scores, respectively, for all of the models in the MSR domain.

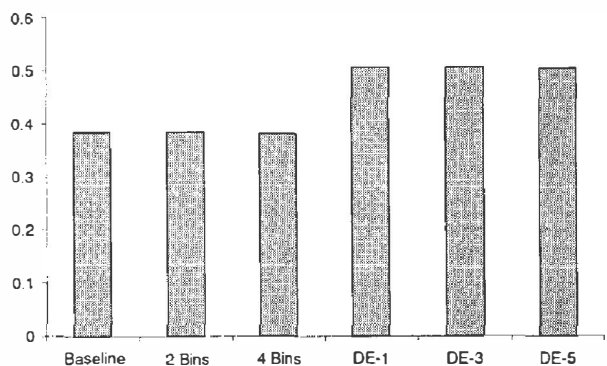

Figure 3: Collaborative filtering scores of the models constructed for the MSR domain.

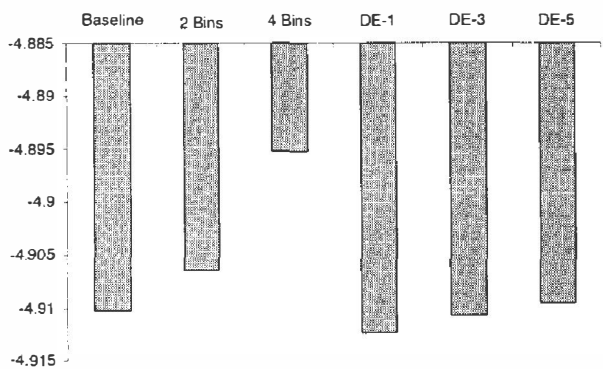

Figure 4: Log-probability scores of the models constructed for the MSNBC domain.

We see that the results are qualitatively almost identical to the MSNBC results. In particular, the data-expansion models are superior for the collaborative-filtering score, but the binning models are superior for the log score. However, binning models do not indicate a steep fall-off in CF accuracy relative to the baseline, as for the MSNBC data set. We hypothesize that typical MSR visitors leave longer page traces than MSNBC users.

## 6 Conclusion

We have presented two techniques for transforming data that allow the collaborative filtering problem to be treated as a time-series prediction task. Both of these techniques allow state-of-the-art collaborative filtering methods to model a richer representation of data when vote sequence information is available. We have evaluated these techniques, using probabilistic decision-tree models, with two data sets for which the order of user votes were known. Results indicate mixed gains for each approach. Binning user data by history length improved log-probability scores with respect to a bag-of-votes model in our test cases, while data expansion to introduce history variables improved the collaborative filtering accuracy score over baseline.